\documentclass{ws-ijmpa}
\usepackage[super,compress]{cite}
\usepackage{graphicx}
\usepackage{braket}
\begin{document}
\markboth{Teruyuki Kitabayashi}{Parametrization of the Yukawa matrix in the scotogenic model and single-zero textures of the neutrino mass matrix}

\def\Journal#1#2#3#4{{#1} {\bf #2}, #3 (#4)}
\def\AHEP{Advances in High Energy Physics.} 
\def\ARNPS{Annu. Rev. Nucl. Part. Sci.} 
\def\AandA{Astron. Astrophys.} 
\def\ANP{Ann. Phys.}
\def\APJ{Astrophys. J.}
\def\APJS{Astrophys. J. Suppl}
\def\COMR{Comptes Rendues}
\def\CPC{Chin. Phys. C}
\def\EPJC{Eur. Phys. J. C}
\def\FrontPhys{Front. Phys.}
\def\IJMPA{Int. J. Mod. Phys. A}
\def\IJMPE{Int. J. Mod. Phys. E}
\def\JCAP{J. Cosmol. Astropart. Phys.}
\def\JHEP{J. High Energy Phys.}
\def\JETPL{JETP. Lett.}
\def\JETPUSSR{JETP (USSR)}
\def\JPG{J. Phys. G} 
\def\MPLA{Mod. Phys. Lett. A}
\def\NIMA{Nucl. Instrum. Meth. A.}
\def\NATU{Nature}
\def\NCA{Nuovo Cimento}
\def\NJP{New. J. Phys.}
\def\NPB{Nucl. Phys. B}
\def\NPBOLD{Nucl. Phys.}
\def\NPBSUPPL{Nucl. Phys. B. Proc. Suppl.}
\def\PLB{{Phys. Lett.} B}
\def\PMCA{PMC Phys. A}
\def\PREP{Phys. Rep.}
\def\PPNP{Prog. Part. Nucl. Phys.}
\def\PLBOLD{Phys. Lett.}
\def\PAN{Phys. Atom. Nucl.}
\def\PRL{Phys. Rev. Lett.}
\def\PRD{Phys. Rev. D}
\def\PRC{Phys. Rev. C}
\def\PR{Phys. Rev.}
\def\PTP{Prog. Theor. Phys.}
\def\PTEP{Prog. Theor. Exp. Phys.}
\def\RMP{Rev. Mod. Phys.}
\def\SJNP{Sov. J. Nucl. Phys.}
\def\SCIENCE{Science}
\def\TNYAS{Trans. New York Acad. Sci.}
\def\ZETP{Zh. Eksp. Teor. Piz.}
\def\ZFPH{Z. fur Physik}
\def\ZPC{Z. Phys. C}

%
\catchline{}{}{}{}{}
%


\title{Parametrization of the Yukawa matrix in the scotogenic model and single-zero textures of the neutrino mass matrix}

\author{Teruyuki Kitabayashi\footnote{
teruyuki@tokai-u.jp}
}

\address{ Department of Physics, Tokai University, 4-1-1 Kitakaname, Hiratsuka, Kanagawa 259-1292, Japan}

\maketitle

\begin{history}
\received{Day Month Year}
\revised{Day Month Year}
\end{history}

\begin{abstract}
As the first topic, we propose a new parametrization of the complex Yukawa matrix in the scotogenic model. The new parametrization is compatible with the particle data group parametrization of the neutrino sector. Some analytical expressions for the neutrino masses with the new parametrization are shown. As the second topic, we consider the phenomenology of the socotogenic model with the one-zero-textures of the neutrino flavor mass matrix. One of the six patterns of the neutrino mass matrix is favorable for the real Yukawa matrix. On the other hand, for the complex Yukawa matrix, five of the six patterns are compatible with observations of the neutrino oscillations, dark matter relic abundance and branching ratio of the $\mu \rightarrow e\gamma$ process.
\end{abstract}

\ccode{PACS numbers:14.60.Pq, 95.35.+d, 98.80.Cq}


\section{Introduction\label{sec:introduction}}
The nature of the dark matter and neutrinos cannot be explained within the standard model of particle physics. The new physics beyond the standard model may provide the hints of these problems. The scotogenic model can simultaneously account for dark matter candidates and the origin of tiny masses of neutrinos \cite{Ma2006PRD}. In this model, neutrino masses are generated by one-loop interactions mediated by a dark matter candidate. One-loop interactions related to dark matter and neutrino mass have been extensively studied in the literature \cite{Ma1998PRL,Kubo2006PLB,Hambye2007PRD,Farzan2009PRD,Farzan2010MPLA,Farzan2011IJMPA,Kanemura2011PRD,Schmidt2012PRD,Faezan2012PRD,Aoki2012PRD,Hehn2012PLB,Bhupal2012PRD,Bhupal2013PRD,Law2013JHEP,Kanemura2013PLB,Hirsch2013JHEP,Restrepo2013JHEP,Ho2013PRD,Lindner2014PRD,Okada2014PRD89,Okada2014PRD90,Brdar2014PLB,Toma2014JHEP,Ho2014PRD,Faisel2014PRD,Vicente2015JHEP,Borah2015PRD,Wang2015PRD,Fraser2016PRD,Adhikari2016PLB,Ma2016PLB,Arhrib2016JCAP,Okada2016PRD,Ahriche2016PLB,Lu2016JCAP,Cai2016JHEP,Suematsu2009PRD,Suematsu2010PRD,Ibarra2016PRD,Lindner2016PRD,Das2017PRD,Singirala2017CPC,Kitabayashi2017IJMPA,AbadaJHEP2018,Rojas2018arXiv,Baumholzer2018JHEP,Ahriche2018PRD,Hugle2018PRD,Kitabayashi2018PRD,Borah2018arXiv,Han2019arXiv,Reig2019PLB}.  

On the other hand, there have been various discussions on flavor neutrino mass matrices with zero elements \cite{Ludl2014JHEP}. The origin of such texture zeros was discussed in Refs.\cite{Berger2001PRD,Low2004PRD,Low2005PRD,Grimus2004EPJC,Xing2009PLB,Dev2011PLB,Araki2012JHEP,Felipe2014NPB,Grimus2005JPG}. In particular, texture of the flavor neutrino mass matrix with single-zero element is called one-zero-textures. There are six patterns for the one-zero-textures of the flavor neutrino mass matrix, which are usually denoted by ${\rm G_1}$, ${\rm G_2}$, $\cdots$, ${\rm G_6}$. The phenomenology of the one-zero-textures was studied, for examples, in Refs.\cite{Xing2004PRD,Lashin2012PRD,Deepthi2012EPJC,Gautam2015PRD}. Also, the experimental potential of probing the texture-zero models has been discussed. For example, see the Ref.\cite{Bora2017PRD} for the possibility of probing different texture-zero flavor mass matrices at DUNE. 

Recently, the author has reported a relation of the scotogenic model with the one-zero-textures of the neutrino flavor mass matrix for the real Yukawa matrix \cite{Kitabayashi2018PRD}. If the elements of the Yukawa matrix in the scotogenic model are real, one of six pattern in the one-zero-textures, ${\rm G_3}$, is favorable. Three patterns (${\rm G_1}, {\rm G_4}$ and ${\rm G_6}$) are impossible in the scotogenic model with the real Yukawa matrix. Two patterns (${\rm G_2}$ and ${\rm G_5}$) are not favorable within the scotogenic model with the real Yukawa matrix because the predicted neutrino oscillation data, the relic abundance of dark matter, and the upper limit of the branching ratio of the $\mu \rightarrow e \gamma$ process should be unlikely or be out of the range of the observed data. 

In this paper, we enlarge the previous argument \cite{Kitabayashi2018PRD} on the relation of the scotogenic model with the one-zero-textures of the neutrino flavor mass matrix to include the complex elements in the Yukawa matrix. There are the following two main topics in this paper:
\begin{itemize}
\item We propose a new parametrization of the complex Yukawa matrix in the scotogenic model. The new parametrization is compatible with the particle data group parametrization of the neutrino sector \cite{PDG}. For the phenomenology of the scotogenic model with the PDG parametrization, the parametrization of the Yukawa matrix which is proposed in this paper may be useful. 
\item In the previous study \cite{Kitabayashi2018PRD}, all elements of the Yukawa matrix are real, so that three patterns (${\rm G_1}, {\rm G_4}$ and ${\rm G_6}$) in the one-zero-textures are impossible in the scotogenic model. If we include the complex Yukawa matrix elements, or equivalently CP-violating phases, the results may have been different. Some analysis of this topic will be found in this paper.
\end{itemize}

This paper is organized as follows.  In Sec.\ref{sec:scotogenic}, a brief review of the scotogenic model is provided. In Sec.\ref{sec:yukawa}, we propose a new parametrization of the complex Yukawa matrix and show some analytical expressions for the neutrino masses in the scotogenic model with the new parametrization. In Sec.\ref{sec:onezero}, some phenomenological study of the scotogenic model with the complex Yukawa matrix in the one-zero-textures of the neutrino flavor neutrino mass matrix scheme will be shown.  Finally, Sec.\ref{sec:summary} is devoted to summary.

\section{Scotogenic model\label{sec:scotogenic}}
We show a brief review of the scotogenic model \cite{Ma2006PRD}. The scotogenic model has three extra Majorana $SU(2)_L$ singlets $N_k$ $(k=1,2,3)$ and one new scalar $SU(2)_L$ doublet $ \eta=(\eta^+,\eta^0)$. $N_k$ and $\eta$ are odd under $Z_2$ symmetry while other fields are even under $Z_2$ symmetry. The Lagrangian of the scotogenic model contains new terms for the new fields,
\begin{eqnarray}
\mathcal{L} \supset Y_{\alpha k} (\bar{\nu}_{\alpha L} \eta^0 - \bar{\ell}_{\alpha L} \eta^+) N_k + \frac{1}{2}M_k \bar{N}_k N^C_k + H.c.,
\label{Eq:L_yukawa}
\end{eqnarray}
and the scalar potential of the model contains the  quartic scalar interaction
\begin{eqnarray}
V \supset \frac{1}{2}\lambda (\Phi^\dagger \eta)^2 + H.c.,
\end{eqnarray}
where $L_\alpha=(\nu_\alpha, \ell_\alpha)$ is the left-handed lepton doublet and $\Phi=(\phi^+, \phi^0)$ is the Higgs doublet in the standard model. The elements of the flavor neutrino mass matrix
\begin{eqnarray}
M_\nu=\left( 
\begin{array}{*{20}{c}}
M_{ee} & M_{e\mu} & M_{e\tau} \\
- & M_{\mu\mu} &M_{\mu\tau}\\
- & - & M_{\tau\tau} \\
\end{array}
\right),
\label{Eq:Mnu}
\end{eqnarray}
where the symbol ``$-$" denotes a symmetric partner, are obtained as

\begin{eqnarray}
M_{\alpha\beta} = \sum_{k=1}^3 Y_{\alpha k}Y_{\beta k} \Lambda_k,
\label{Eq:M_alpha_beta}
\end{eqnarray}
where
\begin{eqnarray}
\Lambda_k &=&  \frac{\lambda v^2}{16\pi^2}\frac{M_k}{m^2_0-M^2_k}\left(1-\frac{M^2_k}{m^2_0-M^2_k}\ln\frac{m_0^2}{M^2_k} \right),
\label{Eq:Lambda_k} \\
m_0^2 &=& \frac{1}{2}(m_R^2+m_I^2),
\label{Eq:LambdaK}
\end{eqnarray}
and $v$, $m_R$, and $m_I$ denote the vacuum expectation value of the Higgs field, and the masses of $\sqrt{2} {\rm Re}[\eta^0]$ and $\sqrt{2} {\rm Im}[\eta^0]$, respectively. 

The scotogenic model predicts the existence of particle dark matter. The lightest $Z_2$ odd particle is stable in the particle spectrum. This lightest $Z_2$ odd particle becomes a dark matter candidate. We assume that the lightest Majorana singlet fermion, $N_1$, becomes the dark matter and $N_1$ is considered to be almost degenerate with the next to lightest Majorana singlet fermion $N_2$, $M_1 \lesssim M_2 < M_3$. In this case, the (co)annihilation cross section times the relative velocity of annihilation particles $v_{\rm rel}$ is given by  \cite{Griest1991PRD,Kubo2006PLB,Suematsu2009PRD,Suematsu2010PRD,Schmidt2012PRD}
\begin{eqnarray}
\sigma_{ij} |v_{\rm rel}|= a_{ij} + b_{ij} v_{\rm rel}^2,
\end{eqnarray}
with
\begin{eqnarray}
a_{ij}&=& \frac{1}{8\pi}\frac{M_1^2}{(M_1^2+m_0^2)^2} \sum_{\alpha\beta}\left|Y_{\alpha i} Y_{\beta j}^* - Y_{\alpha j}^* Y_{\beta i}\right|^2, \\
b_{ij}&=&\frac{m_0^4-3m_0^2M_1^2-M_1^4}{3(M_1^2+m_0^2)^2}a_{ij}   +  \frac{1}{12\pi}\frac{M_1^2(M_1^4+m_0^4)}{(M_1^2+m_0^2)^4}\sum_{\alpha\beta}\left|Y_{\alpha i} Y_{\alpha j}^* Y_{\beta i} Y_{\beta j}^*\right|, \nonumber
\label{Eq:a_b}
\end{eqnarray}
where $\sigma_{ij}$ $(i,j=1,2)$ is annihilation cross section for $N_i N_j \rightarrow \bar{f}f$. The effective cross section $\sigma_{\rm eff}$ is obtained as
\begin{eqnarray}
\sigma_{\rm eff} = \frac{g_1^2}{g_{\rm eff}^2}\sigma_{11} + \frac{2g_1g_2}{g_{\rm eff}^2}\sigma_{12} (1+\Delta M)^{3/2}e^{-\Delta M \cdot x} + \frac{g_2^2}{g_{\rm eff}^2}\sigma_{22} (1+\Delta M)^3 e^{-2\Delta M \cdot x},
\end{eqnarray}
where  $\Delta M = (M_2-M_1)/M_1$ depicts the mass splitting ratio of the degenerate singlet fermions, $x = M_1/T$ denotes the ratio of the mass of lightest singlet fermion to the temperature $T$ and $g_1$ and $g_2$ are the number of degrees of freedom of $N_1$ and $N_2$, respectively, and
\begin{eqnarray}
g_{\rm eff}&=&g_1+g_2 (1+\Delta M)^{3/2}e^{-\Delta M \cdot x}.
\end{eqnarray}

Since $N_1$ is considered almost degenerate with $N_2$, we have $\Delta M \simeq 0$ and obtain
\begin{eqnarray}
\sigma_{\rm eff} |v_{\rm rel}|= a_{\rm eff} + b_{\rm eff} v_{\rm rel}^2,
\end{eqnarray}
where
\begin{eqnarray}
a_{\rm eff}= \frac{a_{11}}{4}+\frac{a_{12}}{2}+\frac{a_{22}}{4},  \quad
b_{\rm eff}= \frac{b_{11}}{4}+\frac{b_{12}}{2}+\frac{b_{22}}{4}.
\label{Eq:aeff_beff}
\end{eqnarray}

The thermally averaged cross section can be written as $\langle \sigma_{\rm eff}|v_{\rm rel}| \rangle = a_{\rm eff} + 6b_{\rm eff}/x$ and the relic abundance of cold dark matter is estimated to be
\begin{eqnarray}
\Omega h^2 = \frac{1.07\times 10^9 x_f}{g_\ast^{1/2} m_{\rm pl}({\rm GeV}) (a_{\rm eff}+3b_{\rm eff}/x_f )},
\end{eqnarray}
where $m_{\rm pl}=1.22\times 10^{19} {\rm GeV}$, $g_{\ast} = 106.75$ and
\begin{eqnarray}
x_f = \ln \frac{0.038 g_{\rm eff} m_{\rm pl} M_1 \langle \sigma_{\rm eff} |v_{\rm rel}| \rangle}{g_\ast^{1/2} x_f^{1/2} }.
\end{eqnarray}

In the scotogenic model, flavor-violating processes such as $\mu \rightarrow e \gamma$ are induced at the one-loop level. The branching ratio of $\mu \rightarrow e \gamma$  is given by \cite{Kubo2006PLB}
\begin{eqnarray}
{\rm Br}(\mu \rightarrow e \gamma)=\frac{3\alpha_{\rm em}}{64\pi(G_Fm_0^2)^2}\left| \sum_{k=1}^3 Y_{\mu k}Y_{e k}^* F \left( \frac{M_k}{m_0}\right) \right|^2,
\end{eqnarray}
where $\alpha_{\rm em}$ denotes the fine-structure constant, $G_F$ denotes the Fermi coupling constant and  $F(x)$ is defined by
\begin{eqnarray}
F(x)=\frac{1-6x^2+3x^4+2x^6-6x^4 \ln x^2}{6(1-x^2)^4}.
\end{eqnarray}
%

\section{Parametrization of Yukawa matrix\label{sec:yukawa}}
\subsection{Yukawa matrix}
In order to obtain any phenomenological prediction in the scotogenic model, the elements of the Yukawa matrix 
\begin{eqnarray}
Y=\left( 
\begin{array}{*{20}{c}}
Y_{e1} & Y_{e2} & Y_{e3} \\
Y_{\mu 1} & Y_{\mu 2} & Y_{\mu 3} \\
Y_{\tau 1} & Y_{\tau 2} & Y_{\tau 3} \\
\end{array}
\right),
\label{Eq:Y}
\end{eqnarray}
should be determined. This matrix is closely connected with the neutrino sector. 

There are several ways for parametrization of the Yukawa matrix. For example, Suematsu, et.al. proposed a parametrization of the Yukawa matrix \cite{Suematsu2010PRD} with the assumption of the tribimaximal mixing in the neutrino sector \cite{Harrison2002PLB530,Xing2002PLB,Harrison2002PLB535}. The exact tribimaximal pattern is approximately consistent with the observed solar and atmospheric neutrino mixings; however, it predicts a vanishing reactor neutrino mixing angle. The observed reactor neutrino mixing angle is small but moderately large. 

Although the exact tribimaximal pattern cannot be the correct description of the neutrino sector, the way to determine the Yukawa matrix elements for the exact tribimaximal pattern \cite{Suematsu2010PRD} is still useful. For example, using the method in Ref. \cite{Suematsu2010PRD}, Singirala proposed the following parametrization \cite{Singirala2017CPC} 
\begin{eqnarray}
Y=\left( 
\begin{array}{*{20}{c}}
 Y_{e1} &  Y_{e2} & Y_{e3} \\
-0.68 Y_{e1}& Y_{e2} & 3.56 Y_{e3} \\
0.31Y_{e1} & -Y_{e2} & 4.55Y_{e3} \\
\end{array}
\right),
\label{Eq:Y_Singirala}
\end{eqnarray}
for an modified tribimaximal mixing \cite{Sruthilaya2015NJP} (see also \cite{Rahat2018PRD})
\begin{eqnarray}
U=U_{\rm MTB}=\left( {\begin{array}{*{20}{c}}
\cos\theta & \sin\theta & 0\\
-\frac{\sin\theta}{\sqrt{2}} & \frac{\cos\theta}{\sqrt{2}} & \frac{1}{\sqrt{2}}\\
\frac{\sin\theta}{\sqrt{2}} & -\frac{\cos\theta}{\sqrt{2}} & \frac{1}{\sqrt{2}}\\
\end{array}} \right) 
 \left( 
\begin{array}{*{20}{c}}
\cos\varphi&0&e^{-i\zeta}\sin\varphi \\
0&1 &0\\
-e^{i\zeta}\sin\varphi &0&\cos\varphi 
\end{array}
 \right),
 \label{Eq:UMTB}
\end{eqnarray}
with $\theta=35^\circ$, $\varphi=12^\circ$ and $\zeta=0$. The corresponding three neutrino mixing angles $\theta_{12}$, $\theta_{23}$, $\theta_{13}$, CP-violating Dirac phase $\delta$ and two Majorana phases $\alpha_2,\alpha_3$ in the particle data group (PDG) parametrization \cite{PDG} are $\theta_{12}=35.60^\circ$, $\theta_{23}=38.05^\circ$, $\theta_{13}=9.80^\circ$ and $\delta=\alpha_2=\alpha_3=0$.

In the next subsection, we propose a new parametrization of the Yukawa matrix in the scotogenic model. We employ the similar strategy in \cite{Suematsu2010PRD} as well as \cite{Singirala2017CPC} to determine the elements of the Yukawa matrix; however, we do not take any assumption in the mixing of the neutrino sector such as $U=U_{\rm MTB}$. The PDG parametrization of the neutrino mixing matrix: $U=U_{\rm PDG}=U_0(\theta_{12}, \theta_{23}, \theta_{13}, \delta) P(\alpha_2, \alpha_3)$ is employed in the new parametrization of the Yukawa matrix. A Yukawa matrix with PDG parametrization have been already proposed in terms of $Y_{\tau 1}$, $Y_{\tau 2}$ and $Y_{\tau 3}$ by Ho and Tandean with $P={\rm diag.}(e^{i\alpha_2/2}, e^{i\alpha_3/2},1)$ \cite{Ho2014PRD}. In this paper, a new PDG compatible Yukawa matrix parametrization will be proposed in terms of $Y_{e 1}$, $Y_{e 2}$ and $Y_{e 3}$ with $P={\rm diag.}(1,e^{i\alpha_2/2}, e^{i\alpha_3/2})$. 
 
It must be emphasized that the parametrization of the Yukawa matrix by Ho and Tandean \cite{Ho2014PRD} is valuable parametrization. Moreover, the commonly used Casas-Ibarra parametrization \cite{Casas2001NPB,Ibarra2016PRD} is powerful to determine the numerical magnitude of the Yukawa matrix elements \cite{Toma2014JHEP,Vicente2015JHEP}. Of course, the numerical determination of the Yukawa matrix with some assumptions such as $U=U_{\rm MTB}$ is worth way. The aim of the next subsection is not denial of these excellent parametrizations of the Yukawa matrix but proposing a new parametrization. For the phenomenology of the scotogenic model in terms of $Y_{e 1}$, $Y_{e 2}$ and $Y_{e 3}$ with the PDG parametrization with $P={\rm diag.}(1,e^{i\alpha_2/2}, e^{i\alpha_3/2})$, the parametrization of the Yukawa matrix which is proposed in this paper may be useful.   

\subsection{Parametrization}
We propose a new parametrization of the Yukawa matrix in this subsection. The PDG parametrization of the neutrino mixing matrix is given as \cite{PDG}
\begin{eqnarray}
U_{\rm PDG}&=&\left( {\begin{array}{*{20}{c}}
c_{12}c_{13} & s_{12}c_{13} & s_{13}e^{- i\delta}\\
- s_{12}c_{23} - c_{12}s_{23}s_{13}e^{i\delta} & c_{12}c_{23} - s_{12}s_{23}s_{13}e^{i\delta} & s_{23}c_{13}\\
s_{12}s_{23} - c_{12}c_{23}s_{13}e^{i\delta} & - c_{12}s_{23} - s_{12}c_{23}s_{13}e^{i\delta} & c_{23}c_{13}
\end{array}} \right) \nonumber \\
&& \times
\left( {\begin{array}{*{20}{c}}
1 & 0 & 0\\
0& e^{i\alpha_2/2}&0\\
0&0&e^{i\alpha_3/2}
\end{array}} \right),
\label{Eq:U_PDG}
\end{eqnarray}
where $c_{ij}=\cos\theta_{ij}$, $s_{ij}=\sin\theta_{ij}$  ($i,j$=1,2,3). Because the relation
\begin{eqnarray}
U_{\rm PDG}^T M_\nu U_{\rm PDG} = {\rm diag}.(m_1,m_2,m_3)
\label{Eq:UTMU},
\end{eqnarray}
with Eqs.(\ref{Eq:Mnu}), (\ref{Eq:M_alpha_beta}) and (\ref{Eq:U_PDG}) has to be satisfied, we obtain 
\begin{eqnarray}
M^{\rm diag}_{11} &=& \sum_{k=1}^3 \Large\{ \Lambda_k(s_{12}(-Y_{\mu k} c_{23}+Y_{\tau k}s_{23}) \nonumber \\
&& +c_{12}(Y_{e k}c_{13}-e^{i\delta}s_{13}(Y_{\tau k}c_{23}+Y_{\mu k}s_{23})))^2 \Large\}= m_1,
\label{Eq:m11=m1}
\end{eqnarray}
\begin{eqnarray}
M^{\rm diag}_{12} &=& \sum_{k=1}^3 \Large\{ e^{i\alpha_2/2}\Lambda_k(-c_{12}^2(Y_{\mu k} c_{23}-Y_{\tau k}s_{23})(-Y_{ek}c_{13}+e^{i\delta}s_{13}(Y_{\tau k}c_{23}+Y_{\mu k}s_{23})))
\nonumber \\
&& \quad +s_{12}^2 (Y_{\mu k}c_{23}-Y_{\tau k}s_{23}) (-Y_{ek}c_{13}+e^{i\delta}s_{13}(Y_{\tau k}c_{23}+Y_{\mu k}s_{23}))
\nonumber \\
&& \quad + c_{12}s_{12}(Y_{ek}^2 c_{13}^2 - Y_{\mu k}^2c_{23}^2 - Y_{\tau k}^2s_{23}^2 - e^{i\delta} Y_{ek} (Y_{\tau k}c_{23}+Y_{\mu k}s_{23})\sin 2\theta_{13} \nonumber \\
&& \quad + e^{2i\delta}s_{13}^2 (Y_{\tau k}c_{23}+Y_{\mu k}s_{23})^2 + Y_{\mu k}Y_{\tau k}\sin 2\theta_{23})\Large\}
= 0,
\label{Eq:m12=0}
\end{eqnarray}
\begin{eqnarray}
M^{\rm diag}_{13} &=& \sum_{k=1}^3 \Large\{ e^{i(\alpha_3-2\delta)/2}\Lambda_k(Y_{e k} s_{13}+e^{i\delta}c_{13}(Y_{\tau k}c_{23}+Y_{\mu k}s_{23}))
\nonumber \\
&& \quad \times (s_{12} (-Y_{\mu k}c_{23}+Y_{\tau k}s_{23}) \nonumber \\
&& \qquad +c_{12} (Y_{ek}c_{13}-e^{i\delta}s_{13}(Y_{\tau k}c_{23}+Y_{\mu k}s_{23})))\Large\}
= 0,
\label{Eq:m13=0}
\end{eqnarray}
\begin{eqnarray}
M^{\rm diag}_{22} &=& \sum_{k=1}^3 \Large\{ e^{i\alpha_2} \Lambda_k(c_{12}(Y_{\mu k} c_{23}-Y_{\tau k}s_{23}) \nonumber \\
&& +s_{12}(Y_{e k}c_{13}-e^{i\delta}s_{13}(Y_{\tau k}c_{23}+Y_{\mu k}s_{23})))^2 \Large\}= m_2,
\label{Eq:m22=m2}
\end{eqnarray}
\begin{eqnarray}
M^{\rm diag}_{23} &=& \sum_{k=1}^3 \Large\{ e^{i(\alpha_2+\alpha_3-2\delta)/2}\Lambda_k(Y_{e k} s_{13}+e^{i\delta}c_{13}(Y_{\tau k}c_{23}+Y_{\mu k}s_{23})) \\
&& \quad \times (c_{12} (Y_{\mu k}c_{23}-Y_{\tau k}s_{23})+s_{12} (Y_{ek}c_{13}-e^{i\delta}s_{13}(Y_{\tau k}c_{23}+Y_{\mu k}s_{23})))\Large\}
= 0, \nonumber 
\label{Eq:m23=0}
\end{eqnarray}
and
\begin{eqnarray}
M^{\rm diag}_{33} = \sum_{k=1}^3 \Large\{ e^{i(\alpha_3-2\delta)} \Lambda_k(Y_{e k} s_{13}+e^{i\delta}c_{13}(Y_{\tau k}c_{23}+Y_{\mu k}s_{23}))^2 \Large\}= m_3,
\label{Eq:m33=m3}
\end{eqnarray}
where $m_1$, $m_2$ and $m_3$ denote the neutrino mass eigenvalues. 

The Eqs.(\ref{Eq:m12=0}), (\ref{Eq:m13=0}) and (\ref{Eq:m23=0}) yield the following Yukawa matrix
\begin{eqnarray}
Y=\left( 
\begin{array}{*{20}{c}}
Y_{e1} & Y_{e2} & Y_{e3} \\
a_1 Y_{e1} & a_3 Y_{e2} &  a_5 Y_{e3} \\
a_2 Y_{e1} & a_4Y_{e2} &  a_6Y_{e3} \\
\end{array}
\right),
\label{Eq:Y_Ye1Ye2Ye3}
\end{eqnarray}
where
\begin{eqnarray}
a_1 &=& -\frac{c_{23}t_{12}}{c_{13}}-e^{-i\delta}s_{23}t_{13}, \quad
a_2 = \frac{s_{23}t_{12}}{c_{13}}-e^{-i\delta}c_{23}t_{13}, \nonumber \\
a_3 &=& \frac{c_{23}}{t_{12}c_{13}}-e^{-i\delta}s_{23}t_{13}, \quad
a_4 = -\frac{s_{23}}{t_{12}c_{13}}-e^{-i\delta}c_{23}t_{13}, \nonumber \\
a_5 &=& e^{-i\delta}\frac{s_{23}}{t_{13}}, \quad a_6 = e^{-i\delta}\frac{c_{23}}{t_{13}},
\label{Eq:a1a2a3a4a5a6}
\end{eqnarray}
and $t_{ij}=\tan\theta_{ij}$  ($i,j$=1,2,3). The Eq.(\ref{Eq:Y_Ye1Ye2Ye3}) with Eq.(\ref{Eq:a1a2a3a4a5a6}) is the new parametrization of the Yukawa matrix which we proposed in this paper. This parametrization of the Yukawa matrix elements is also relevant for extended scotogenic models if the flavor neutrino masses are expressed as $M_{\alpha\beta} = \sum_{k=1}^3 Y_{\alpha k}Y_{\beta k} \Lambda_k$. 

Now, we check the reproducibility of the Singirala's result in Eq.(\ref{Eq:Y_Singirala}). If we take 
\begin{eqnarray}
&& \theta_{12}=35.60^\circ, \quad \theta_{23}=38.05^\circ, \quad \theta_{13}=9.80^\circ, \nonumber \\
&& \delta=\alpha_2=\alpha_3=0^\circ,
\end{eqnarray}
to compere the coefficients of the Yukawa matrix elements with Singirala's numerical result, we obtain 
\begin{eqnarray}
&& a_1 = -0.679, \quad a_3 = 1.01, \quad a_5 = 3.57, \nonumber \\
&& a_2 = 0.312, \quad a_4 = -1.01, \quad a_6 = 4.56, 
\end{eqnarray}
and these values are consistent with Eq.(\ref{Eq:Y_Singirala}).

Some analytical expressions for the neutrino sector in the scotogenic model with the new parametrization of the Yukawa matrix in Eqs.(\ref{Eq:Y_Ye1Ye2Ye3}) and (\ref{Eq:a1a2a3a4a5a6}) are obtained as follows.

The Eqs.(\ref{Eq:m11=m1}), (\ref{Eq:m22=m2}) and (\ref{Eq:m33=m3}) yield the following neutrino mass eigenvalues
\begin{eqnarray}
m_i = b_i \Lambda_i Y_{ei}^2,
\label{Eq:m1m2m3}
\end{eqnarray}
where
\begin{eqnarray}
b_1 = \frac{1}{c_{12}^2c_{13}^2}, \quad 
b_2 = \frac{e^{i\alpha_2}}{s_{12}^2c_{13}^2}, \quad
b_3 = \frac{e^{i(\alpha_3-2\delta)}}{s_{13}^2}.
\label{Eq:b1b2b3}
\end{eqnarray}

Although the neutrino mass ordering (either the normal mass ordering or the inverted mass ordering) is not determined, a global analysis shows that the preference for the normal mass ordering is mostly due to neutrino oscillation measurements \cite{Salas2018arXiv,Salas2018PLB}. We assume the normal mass ordering (NO) for the neutrinos. In this case, the squared mass differences of the neutrinos are given by 
\begin{eqnarray}
\Delta m_{21}^2 = m_2^2-m_1^2, \quad \Delta m_{31}^2 = m_3^2-m_1^2, 
\label{Eq:DeltaM_NO}
\end{eqnarray}
and we obtain the relations
\begin{eqnarray}
Y_{e2}^2 &=& \frac{\sigma_2}{b_2\Lambda_2}\sqrt{\Delta m_{21}^2+ b_1^2 \Lambda_1^2 Y_{e1}^4}, \nonumber \\
Y_{e3}^2 &=& \frac{\sigma_3}{b_3\Lambda_3}\sqrt{\Delta m_{31}^2+ b_1^2 \Lambda_1^2 Y_{e1}^4},
\label{Eq:Y2Y3}
\end{eqnarray}
where $\sigma_{2,3}=\pm 1$. We take $\sigma_{2,3}= 1$ \cite{Singirala2017CPC}. Since of the relation $\Lambda_k=f(\lambda, m_0, M_k)$, the elements of the Yukawa matrix can be determined  as
\begin{eqnarray} 
Y_{\alpha k} = f(\theta_{ij}, \delta, \alpha_i,\Delta m_{ij}^2; Y_{e1}; \lambda, m_0, M_k),
\end{eqnarray}
where $\theta_{ij}, \delta, \alpha_i,\Delta m_{ij}^2$ are neutrino sector parameters, $\lambda, m_0, M_k$ are dark sector parameter, and $Y_{e1}$ bridges these two sectors.

The $ee$-element of the flavor neutrino mass matrix $M_\nu$ can be written as 
\begin{eqnarray}
M_{ee} &=&  Y_{e1}^2\Lambda_1 + Y_{e2}^2\Lambda_2 + Y_{e3}^2\Lambda_3 \nonumber \\
 &=&  Y_{e1}^2\Lambda_1 + \frac{\sigma_2}{b_2}\sqrt{\Delta m_{21}^2+ b_1^2 \Lambda_1^2 Y_{e1}^4}  +  \frac{\sigma_3}{b_3}\sqrt{\Delta m_{31}^2+ b_1^2 \Lambda_1^2 Y_{e1}^4}.
 \label{Eq:Mee}
\end{eqnarray}
Similarly, we obtain
\begin{eqnarray}
M_{e\mu}  &=&  a_1 Y_{e1}^2\Lambda_1 + \frac{\sigma_2 a_3}{b_2}\sqrt{\Delta m_{21}^2+ b_1^2 \Lambda_1^2 Y_{e1}^4} 
 +  \frac{\sigma_3 a_5}{b_3}\sqrt{\Delta m_{31}^2+ b_1^2 \Lambda_1^2 Y_{e1}^4},
 \label{Eq:Meu}
\end{eqnarray}
\begin{eqnarray}
M_{e\tau}  &=&  a_2 Y_{e1}^2\Lambda_1 + \frac{\sigma_2 a_4}{b_2}\sqrt{\Delta m_{21}^2+ b_1^2 \Lambda_1^2 Y_{e1}^4} 
 +  \frac{\sigma_3 a_6}{b_3}\sqrt{\Delta m_{31}^2+ b_1^2 \Lambda_1^2 Y_{e1}^4},
 \label{Eq:Met}
\end{eqnarray}
\begin{eqnarray}
M_{\mu\mu}  &=&  a_1^2 Y_{e1}^2\Lambda_1 + \frac{\sigma_2 a_3^2}{b_2}\sqrt{\Delta m_{21}^2+ b_1^2 \Lambda_1^2 Y_{e1}^4} 
 +  \frac{\sigma_3 a_5^2}{b_3}\sqrt{\Delta m_{31}^2+ b_1^2 \Lambda_1^2 Y_{e1}^4},
 \label{Eq:Muu}
\end{eqnarray}
\begin{eqnarray}
M_{\mu\tau}  &=&  a_1a_2 Y_{e1}^2\Lambda_1 + \frac{\sigma_2 a_3a_4}{b_2}\sqrt{\Delta m_{21}^2+ b_1^2 \Lambda_1^2 Y_{e1}^4} 
  +  \frac{\sigma_3 a_5 a_6}{b_3}\sqrt{\Delta m_{31}^2+ b_1^2 \Lambda_1^2 Y_{e1}^4}, \nonumber \\
\label{Eq:Mut}
\end{eqnarray}
and
\begin{eqnarray}
M_{\tau\tau} &=&  a_2^2 Y_{e1}^2\Lambda_1 + \frac{\sigma_2 a_4^2}{b_2}\sqrt{\Delta m_{21}^2+ b_1^2 \Lambda_1^2 Y_{e1}^4}  
 +  \frac{\sigma_3 a_6^2}{b_3}\sqrt{\Delta m_{31}^2+ b_1^2 \Lambda_1^2 Y_{e1}^4}.
 \label{Eq:Mtt}
\end{eqnarray}
Thus, $M_{\alpha\beta}$ is independent of $\Lambda_2$ and $\Lambda_3$ as well as $M_2$ and $M_3$:
\begin{eqnarray} 
M_{\alpha\beta} = f(\theta_{ij}, \delta, \alpha_i,\Delta m_{ij}^2; Y_{e1}; \lambda, m_0, M_1).
\end{eqnarray}
%

\section{One-zero-textures \label{sec:onezero}}
\subsection{Real Yukawa matrix \label{subsec:one_zero_real_yukawa}}
As we addressed in introduction, we assume that the flavor neutrino mass matrix $M_\nu$ has one zero element. There are the following six patterns for the flavor neutrino mass matrix $M_\nu$ in the one-zero-textures:
\begin{eqnarray}
&& {\rm G}_1:
\left( 
\begin{array}{*{20}{c}}
0 & \times & \times \\
- & \times & \times \\
- & - & \times \\
\end{array}
\right),
\quad
 {\rm G}_2:
\left( 
\begin{array}{*{20}{c}}
\times & 0 & \times \\
- & \times & \times \\
- & - & \times \\
\end{array}
\right),
\nonumber \\
&&{\rm G}_3:
\left( 
\begin{array}{*{20}{c}}
\times & \times &0 \\
- & \times & \times \\
- & - & \times \\
\end{array}
\right),
\quad
{\rm G}_4:
\left( 
\begin{array}{*{20}{c}}
\times & \times & \times \\
- & 0 & \times \\
- & - & \times \\
\end{array}
\right),
\nonumber \\
&& {\rm G}_5:
\left( 
\begin{array}{*{20}{c}}
\times & \times & \times \\
- & \times & 0 \\
- & - & \times \\
\end{array}
\right),
\quad 
{\rm G}_6:
\left( 
\begin{array}{*{20}{c}}
\times & \times & \times \\
- & \times & \times \\
- & - &0 \\
\end{array}
\right).
\label{Eq:G1G2G3G4G5G6}
\end{eqnarray}

First, we assume that all elements of the Yukawa matrix are real. Because the present data disfavor CP conservation, the assumption of a real Yukawa coupling matrix is not realistic; however, we start our study without CP violation just as a simple case. In this case, the author has already reported that one of six patterns, ${\rm G_3}$, is favorable for the observed neutrino oscillation data, the relic abundance of dark matter, and the upper limit of the branching ratio of the $\mu \rightarrow e \gamma$ process \cite{Kitabayashi2018PRD}. In this previous study \cite{Kitabayashi2018PRD}, a specific parametrization of the Yukawa matrix with an assumption of the neutrino mixing (modified tribimaximal mixing) was employed \cite{Singirala2017CPC}. In this subsection, we briefly reproduce the results in Ref.\cite{Kitabayashi2018PRD} by using the new parametrization of the Yukawa matrix in Eq.(\ref{Eq:Y_Ye1Ye2Ye3}) with Eq.(\ref{Eq:a1a2a3a4a5a6}).

For the ${\rm G_1}$ pattern, the relation
\begin{eqnarray}
M_{ee} =  Y_{e1}^2\Lambda_1 + Y_{e2}^2\Lambda_2 + Y_{e3}^2\Lambda_3 =0
\label{Eq:MeeG1}
\end{eqnarray}
is required by Eq.(\ref{Eq:M_alpha_beta}). Since $\Lambda_k > 0$ and $Y_{\alpha k}$ is real, Eq.(\ref{Eq:MeeG1}) yields $Y_{ek}=0$. However, the vanishing $Y_{ek}$ yields
\begin{eqnarray}
M_{e\mu} = \sum_{k=1}^3 Y_{ek}Y_{\mu k}\Lambda_k  =0, \nonumber \\
M_{e\tau} = \sum_{k=1}^3 Y_{ek}Y_{\tau k}\Lambda_k  =0,
\end{eqnarray}
as well as 
\begin{eqnarray}
\left( 
\begin{array}{*{20}{c}}
0 & 0 & 0 \\
- & \times & \times \\
- & - & \times \\
\end{array}
\right),
\end{eqnarray}
and the one-zero-texture assumption should be violated. The ${\rm G_1}$ pattern is excluded in the scotogenic model. Similarly, the ${\rm G_4}$ and ${\rm G_6}$ patterns are also excluded. Thus, three patterns (${\rm G_1}, {\rm G_4}$ and ${\rm G_6}$) are impossible in the scotogenic model with the real Yukawa matrix. 

To see that whether the ${\rm G_2}$, ${\rm G_3}$ and ${\rm G_5}$ are consistent with observation or not, we performed numerical calculations with the following input values: 
\begin{itemize}
\item For neutrino sector, we fix the masses of the neutrinos with the best-fit values of the squared mass differences $\Delta m_{ij}^2$ \cite{Esteban2017JHEP}
\begin{eqnarray} 
\Delta m^2_{21} &=& 7.50 \times 10^{-5} {\rm eV}^2, \nonumber \\
\Delta m^2_{31} &=& 2.524 \times 10^{-3} {\rm eV}^2,
\label{Eq:best_fit_deltams}
\end{eqnarray}
and vary the mixing angles in the following $3 \sigma$ region \cite{Esteban2017JHEP}
\begin{eqnarray}
&& \theta_{12}/^\circ = 31.38 - 35.99, \nonumber \\
&& \theta_{23}/^\circ = 38.4 - 52.8, \nonumber \\
&& \theta_{13}/^\circ = 7.99 - 8.90.
\label{Eq:mixing_angle}
\end{eqnarray}
\item For dark sector, we adopt the following standard criteria \cite{Kubo2006PLB,Ibarra2016PRD,Lindner2016PRD}. 1) The quartic coupling satisfies the relation $|\lambda| \ll 1$ for small neutrino masses. 2) Since we assumed that the additional lightest Majorana fermion $N_1$ is dark matter particle, we require $M_1 \le M_2, M_3, m_0$. 3) The mass scale of new fields is a few TeV.  We take 
\begin{eqnarray}
1 \times 10^{-10} \le &\lambda& \le 5 \times 10^{-9}, \nonumber \\
0.5 \le &r_1& \le 0.9, \nonumber \\
1.5 \le &r_3& \le 3.0, \nonumber \\
2 {\rm TeV} \le &m_0& \le 4 {\rm TeV},
\label{Eq:DMparameteter}
\end{eqnarray}
where
\begin{eqnarray}
r_k=\frac{M_k}{m_0}.
\end{eqnarray}
\end{itemize}

The ${\rm G_3}$ pattern is consistent with the observations. For example, the best-fit values of neutrino mixing angles%
\begin{eqnarray}
\theta_{12}/^\circ = 33.56, \quad
\theta_{23}/^\circ = 41.6, \quad
\theta_{13}/^\circ = 8.46,
\label{Eq:mixing_angle_bestfit}
\end{eqnarray}
with 
\begin{eqnarray}
\lambda= 4.63 \times 10^{-9}, \ r_1 = 0.9, \ r_3 = 1.5, \ m_0= 3{\rm TeV},
\label{Eq:benchmark_parameter_set}
\end{eqnarray}
 yield
\begin{eqnarray}
\Omega h^2 = 0.120, \quad {\rm Br}(\mu \rightarrow e\gamma) = 2.50 \times 10^{-13},
\end{eqnarray}
for $G_3$, which are consistent with the observed energy density of the cold dark matter component in the ${\rm \Lambda CDM}$ cosmological model by the Plank Collaboration $\Omega h^2 = 0.120 \pm 0.001$ \cite{Planck} and the measured upper limit of the branching ratio $ {\rm Br}(\mu \rightarrow e\gamma) \le 4.2 \times 10^{-13}$ \cite{MEG2016EPJC}. Although the upper limits of the branching ratio of ${\rm Br}(\tau \rightarrow \mu \gamma) \le 4.4 \times 10^{-8}$ and ${\rm Br}(\tau \rightarrow e \gamma) \le 3.3 \times 10^{-8}$ were also reported \cite{BABAR2010PRL}, we only account for ${\rm Br}(\mu \rightarrow e\gamma)$ since it is the most stringent constraint. 

On the other hand, it turned out that the predicted values of $\Omega h^2$ and $ {\rm Br}(\mu \rightarrow e\gamma)$ for ${\rm G_2}$ seems to be  unlikely with the observed data.  Moreover, ${\rm G_5}$ pattern is excluded from observation. The ${\rm G_2}$ and ${\rm G_5}$ patterns are not favorable for the scotogenic model with the real Yukawa matrix elements. For more detail, see Ref.\cite{Kitabayashi2018PRD}.

\subsection{Complex Yukawa matrix \label{subsec:one_zero_complex_yukawa}}
In the previous subsection, all elements of the Yukawa matrix are real, there is no CP-violating source in the Yukawa sector, so that three patterns (${\rm G_1}, {\rm G_4}$ and ${\rm G_6}$) in the one-zero-textures are impossible in the scotogenic model. If we include CP-violating phases, the results to be different. Some analysis of this topic will be found in this subsection.

\begin{figure}[t]
\begin{center}
\includegraphics[width=6cm]{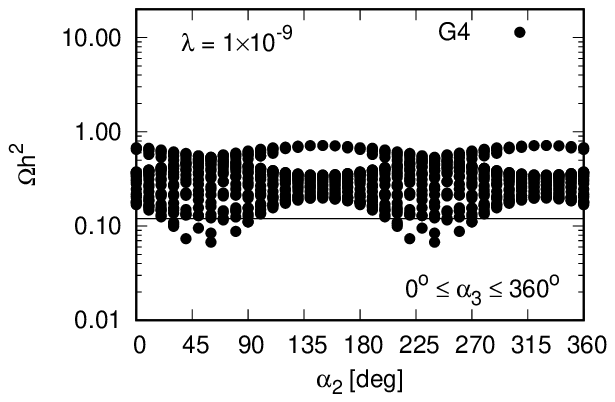}
\includegraphics[width=6cm]{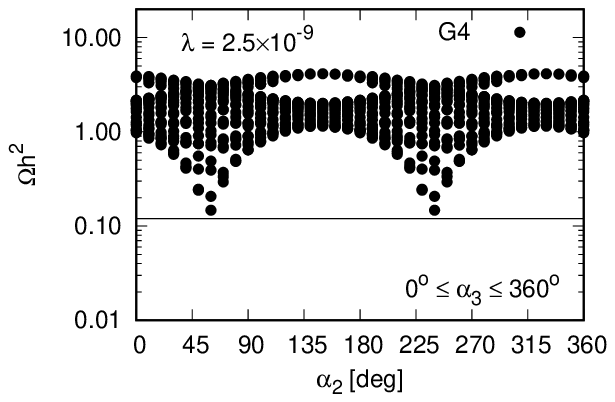}
\includegraphics[width=6cm]{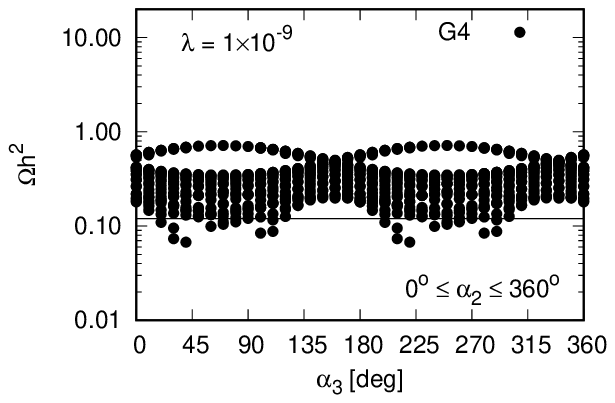}
\includegraphics[width=6cm]{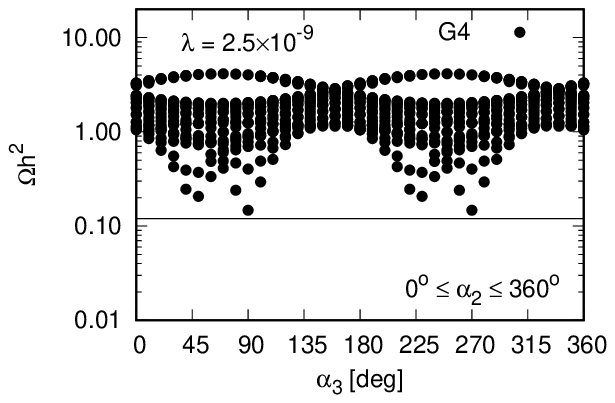}
\caption{The dependence of the relic abundance of dark matter $\Omega h^2$ on the coupling $\lambda$ and Majorana CP phases $\alpha_2, \alpha_3$ for ${\rm Br}(\mu \rightarrow e\gamma) \le  4.2\times 10^{-13}$ in the ${\rm G_4}$ pattern. These panels show $\Omega h^2$ vs $\alpha_2$ (upper two panels) or $\alpha_3$ (lower two panels) for $\lambda = 1 \times 10^{-9}$ (left two panels) or $\lambda = 2.5 \times 10^{-9}$ (right two panels). In all panels, other model parameters are fixed as $r_1=0.9$, $r_3=1.5$ and $m_0=3$ TeV. The horizontal line shows $\Omega h^2 = 0.12$.}
\label{fig:omegah2_lambda_G4}
\end{center}
\end{figure}
\begin{figure}[h]
\begin{center}
\includegraphics[width=6cm]{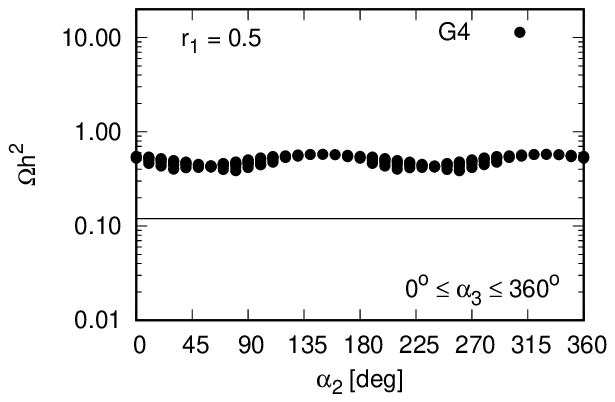}
\includegraphics[width=6cm]{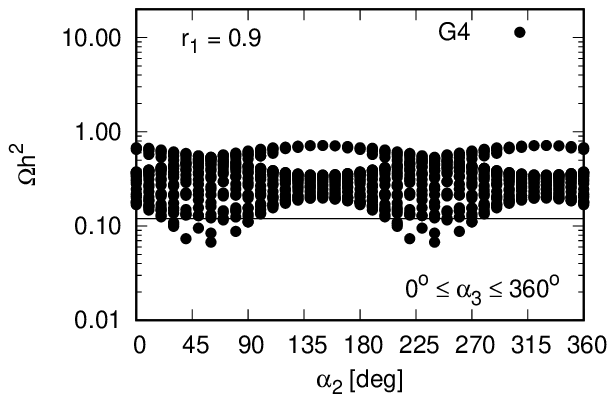}
\includegraphics[width=6cm]{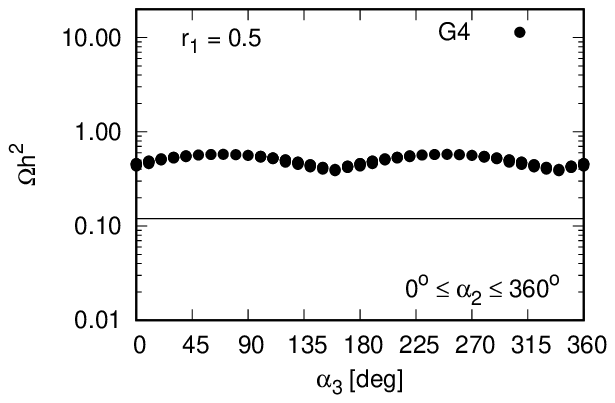}
\includegraphics[width=6cm]{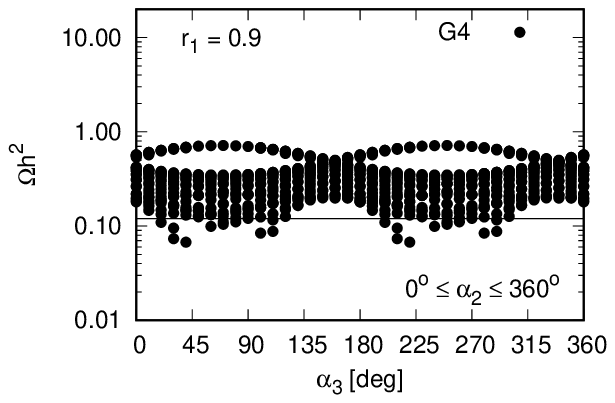}
\caption{The dependence of the relic abundance of dark matter $\Omega h^2$ on the mass ratio $r_1$ and Majorana CP phases $\alpha_2, \alpha_3$ for ${\rm Br}(\mu \rightarrow e\gamma) \le  4.2\times 10^{-13}$ in the ${\rm G_4}$ pattern. These panels show $\Omega h^2$ vs $\alpha_2$ (upper two panels) or $\alpha_3$ (lower two panels) for $r_1=0.5$ (left two panels) or $r_1=0.9$ (right two panels).  In all panels, other model parameters are fixed as  $\lambda = 1 \times 10^{-9}$, $r_3=1.5$ and $m_0=3$ TeV. The horizontal line shows $\Omega h^2 = 0.12$.}
\label{fig:omegah2_r1_G4}
\end{center}
\end{figure}
\begin{figure}[h]
\begin{center}
\includegraphics[width=6cm]{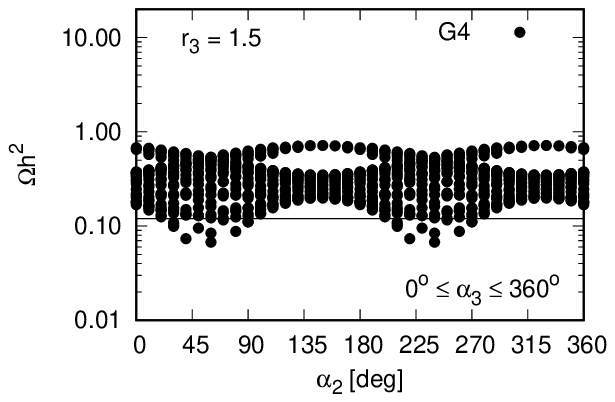}
\includegraphics[width=6cm]{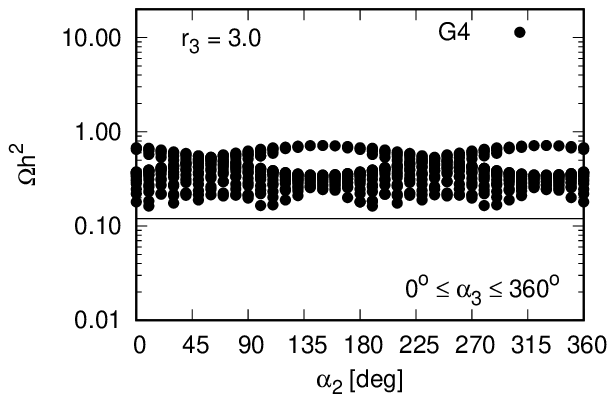}
\includegraphics[width=6cm]{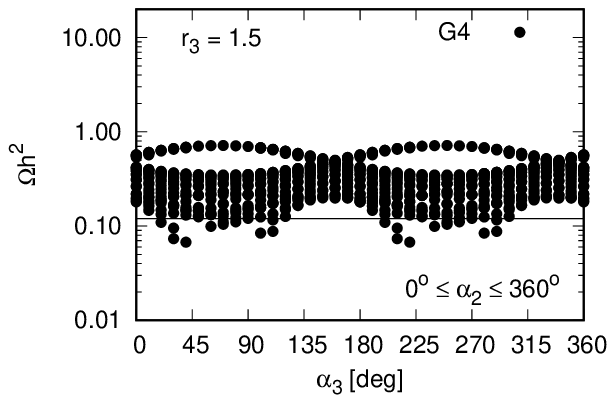}
\includegraphics[width=6cm]{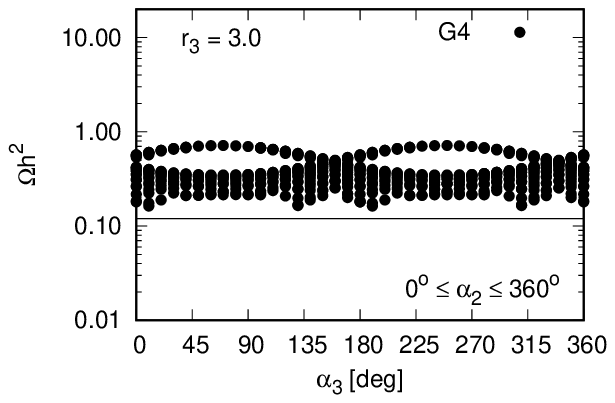}
\caption{The dependence of the relic abundance of dark matter $\Omega h^2$ on the mass ratio $r_3$ and Majorana CP phases $\alpha_2, \alpha_3$ for ${\rm Br}(\mu \rightarrow e\gamma) \le  4.2\times 10^{-13}$ in the ${\rm G_4}$ pattern. These panels show $\Omega h^2$ vs $\alpha_2$ (upper two panels) or $\alpha_3$ (lower two panels) for $r_3=1.5$ (left two panels) or $r_3=3.0$ (right two panels). In all panels, other model parameters are fixed as $\lambda = 1 \times 10^{-9}$, $r_1=0.9$ and $m_0=3$ TeV. The horizontal line shows $\Omega h^2 = 0.12$.}
\label{fig:omegah2_r3_G4}
\end{center}
\end{figure}
\begin{figure}[h]
\begin{center}
\includegraphics[width=6cm]{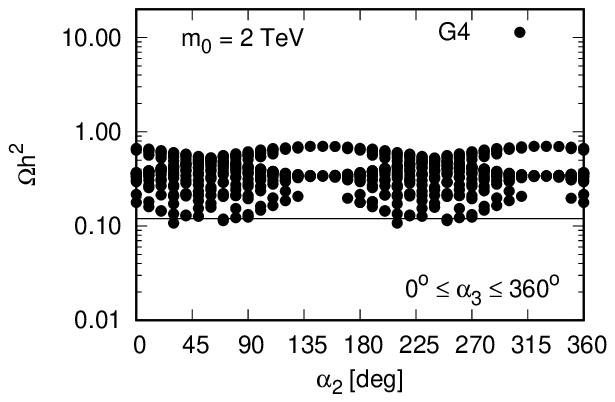}
\includegraphics[width=6cm]{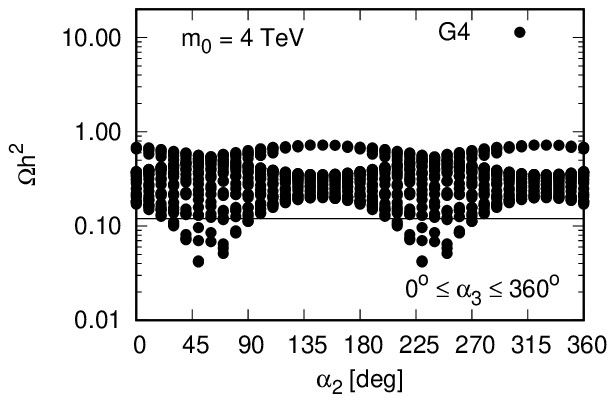}
\includegraphics[width=6cm]{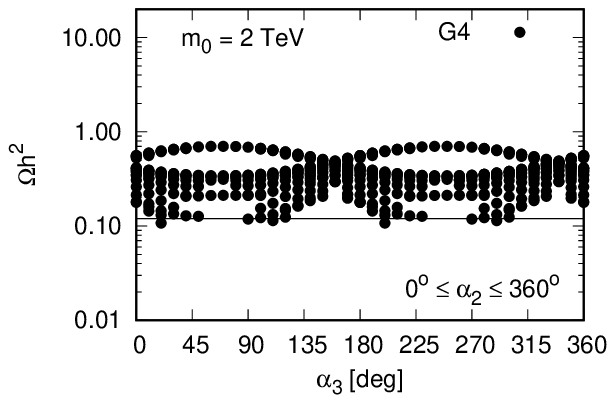}
\includegraphics[width=6cm]{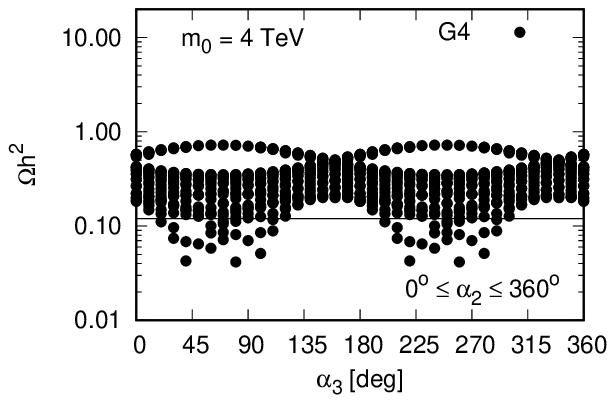}
\caption{The dependence of the relic abundance of dark matter $\Omega h^2$ on the scalar mass  $m_0$ and Majorana CP phases $\alpha_2, \alpha_3$ for ${\rm Br}(\mu \rightarrow e\gamma) \le  4.2\times 10^{-13}$ in the ${\rm G_4}$ pattern. These panels show $\Omega h^2$ vs $\alpha_2$ (upper two panels) or $\alpha_3$ (lower two panels) for $m_0=2$TeV (left two panels) or $m_0=4$TeV (right two panels). In all panels, other model parameters are fixed as $\lambda = 1 \times 10^{-9}$, $r_1=0.9$ and $r_3=1.5$. The horizontal line shows $\Omega h^2 = 0.12$.}
\label{fig:omegah2_m0_G4}
\end{center}
\end{figure}

We performed numerical calculations with the following input values: 
\begin{itemize}
\item For neutrino sector, we fix the masses and mixing of the neutrinos with the best-fit values in Eq.(\ref{Eq:best_fit_deltams}) and Eq.(\ref{Eq:mixing_angle_bestfit}). The coefficients of the Yukawa matrix elements and the mass eigenvalues are estimated to be
\begin{eqnarray}
a_1 &=& -0.502 -0.0988 e^{-i\delta}, \nonumber \\
a_2 &=& 0.445 -0.111 e^{-i\delta}, \nonumber \\
a_3 &=& 1.14-0.0988e^{-i\delta}, \nonumber \\
a_4 &=& -1.01-0.111e^{-i\delta}, \nonumber \\
a_5 &=& 4.46 e^{-i\delta}, \quad a_6 = 5.03 e^{-i\delta}, 
\label{Eq:a1a2a3a4a5a6_bestfit}
\end{eqnarray}
and 
\begin{eqnarray}
b_1 &=& 1.47, \quad b_2 = 3.34 e^{i\alpha_2}, \nonumber \\
b_3 &=& 46.2 e^{i(\alpha_3-2\delta)} .
\label{Eq:b1b2b3_bestfit}
\end{eqnarray}
The Dirac CP phase is fixed in the following best-fit values \cite{Esteban2017JHEP}:
\begin{eqnarray}
\delta = 261^\circ
\end{eqnarray}
and Majorana CP phases are varied as
\begin{eqnarray}
0^\circ \le \alpha_2, \alpha_3 \le 360^\circ.
\end{eqnarray}

\item For dark sector, we adopt the same criteria in the previous subsection.

\end{itemize}

For illustration, 
\begin{eqnarray}
&& \lambda=  1.0 \times 10^{-9}, \ r_1 =0.9 , \ r_3 = 1.5, \ m_0= 3{\rm TeV}, \nonumber \\
&& \delta=  261^\circ, \quad \alpha_2 =34.6^\circ ,\quad \alpha_3= 20.1^\circ, 
\label{Eq:param_comlex}
\end{eqnarray}
yield
\begin{eqnarray}
\Omega h^2 = 0.120, \quad {\rm Br}(\mu \rightarrow e\gamma) =  3.3\times 10^{-13},
\end{eqnarray}
for $G_3$, which are consistent with the observed energy density of the cold dark matter component and the measured upper limit of the branching ratio of $\mu \rightarrow e\gamma$ process.

From more general parameter search, it turned out that ${\rm G_1}$ pattern is unfavorable from observation. On the other hand, remaining five patterns of ${\rm G_2}$, ${\rm G_3}$, $\cdots$, ${\rm G_6}$ are consistent with observation. To see these results, first we show the results from a  parameter search for $G_4$ in FIG.\ref{fig:omegah2_lambda_G4} - FIG.\ref{fig:omegah2_m0_G4}. Figure \ref{fig:omegah2_lambda_G4} shows that the dependence of the relic abundance of dark matter $\Omega h^2$ on the coupling $\lambda$ and Majorana CP phases $\alpha_2, \alpha_3$ for ${\rm Br}(\mu \rightarrow e\gamma) \le  4.2\times 10^{-13}$ in the ${\rm G_4}$ pattern. These panels show $\Omega h^2$ vs $\alpha_2$ (upper two panels) or $\alpha_3$ (lower two panels) for $\lambda = 1 \times 10^{-9}$ (left two panels) or $\lambda = 2.5 \times 10^{-9}$ (right two panels). In all panels, other model parameters are fixed as $r_1=0.9$, $r_3=1.5$ and $m_0=3$ TeV. The horizontal line shows $\Omega h^2 = 0.12$. FIGs.\ref{fig:omegah2_r1_G4}, \ref{fig:omegah2_r3_G4} and \ref{fig:omegah2_m0_G4} are the same as FIG.\ref{fig:omegah2_lambda_G4} but for the dependence of $\Omega h^2$ on $r_1$, $r_3$ and $m_0$. From FIG.\ref{fig:omegah2_lambda_G4} - FIG.\ref{fig:omegah2_m0_G4}, we see the existence of the allowed parameter set \{$\alpha_2, \alpha_3, \lambda, r_1, r_3, m_0$\} for the observed $\Omega h^2$ and ${\rm Br}(\mu \rightarrow e\gamma)$ in ${\rm G_4}$ case. Similar results are obtained for ${\rm G_2}$, ${\rm G_3}$, ${\rm G_5}$ and ${\rm G_6}$ (the dependence of $\Omega h^2$ on $\alpha_2$, $\alpha_3$, $\lambda$, $r_1$, $r_3$ or $m_0$ is most clearly shown in the case of ${\rm G_4}$, so that we chose the case of ${\rm G_4}$ for illustration). Thus we conclude that the five patterns of ${\rm G_2}$, ${\rm G_3}$, ${\rm G_4}$, ${\rm G_5}$ and ${\rm G_6}$ are consistent with observations.

\begin{figure}[t]
\begin{center}
\includegraphics[width=6cm]{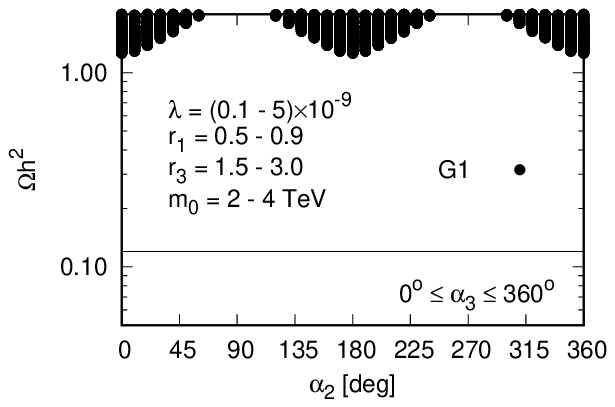}
\includegraphics[width=6cm]{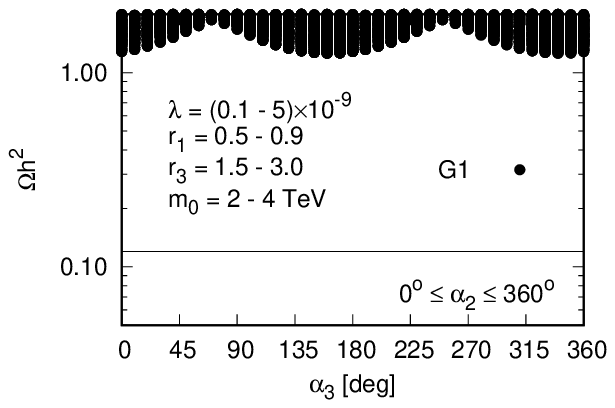}
\caption{The relic abundance of dark matter $\Omega h^2$ vs the Majorana phase $\alpha_2$ (upper panel) or $\alpha_3$ (lower panel) for ${\rm Br}(\mu \rightarrow e\gamma) \le  4.2\times 10^{-13}$ in the ${\rm G_1}$ pattern. The horizontal line shows $\Omega h^2 = 0.12$.}
\label{fig:omegah2_alpha_G1}
\end{center}
\end{figure}

On the contrary, ${\rm G_1}$ pattern is unfavorable with observations. Figure.\ref{fig:omegah2_alpha_G1} shows that the relic abundance of dark matter $\Omega h^2$ vs the Majorana phase $\alpha_2$ (upper panel) or $\alpha_3$ (lower panel) in the ${\rm G_1}$ pattern for the parameter space in Eq.(\ref{Eq:DMparameteter}). The horizontal line shows $\Omega h^2 = 0.12$. The ${\rm G_1}$ pattern is excluded from observations of the neutrino oscillation (best-fit values), dark matter relic abundance and branching ratio of $\mu \rightarrow e\gamma$ in the parameter space in Eq.(\ref{Eq:DMparameteter}). If we expand the model parameter space and/or we allow $3 \sigma$ data of neutrino oscillation experiment instead of best-fit values, some points in ${\rm G_1}$ pattern might become consistent with observations. Even if it is correct, we can say that ${\rm G_1}$ pattern is unfavorable with observations. 

We also estimate the effective neutrino mass for the neutrinoless double beta decay. The neutrinoless double beta decay is allowed if neutrinos are massive Majorana particles \cite{DellOro2016AHEP}. The half-life of the neutrinoless double beta decay is proportional to the effective neutrino mass $m_{\beta\beta}$ which depends on the Dirac and the Majorana CP phases. From numerical calculations, we obtain $m_{\beta\beta} = \mathcal{O}(0.01 - 0.1)$ eV for $\Omega h^2 = 0.120 \pm 0.001$ and  $ {\rm Br}(\mu \rightarrow e\gamma) \le 4.2 \times 10^{-13}$ in ${\rm G_2}$, ${\rm G_3}$, ${\rm G_4}$, ${\rm G_5}$, ${\rm G_6}$ ($m_{\beta\beta}=0$ for ${\rm G_1}$ \cite{Barreiros2018arXiv}). The estimated magnitude of the effective Majorana neutrino mass from experiments is $m_{\beta\beta}$[eV] $\lesssim 0.15 - 2.1$ \cite{DellOro2016AHEP}. In future experiments, a desired sensitivity of $m_{\beta\beta} \lesssim 0.01$ eV will be reached and we may obtain some constraints on the parameters in the scotogenic model from the future neutrinoless double beta decay experiments.

We would like to comment that the discrimination of realistic one-zero textures with complex Yukawa matrix is much less powerful than the case of real couplings. This suggests that it may be useful to study two-zero textures \cite{Kitabayashi2017IJMPA,Dev2014PRD,Meloni2014PRD,twoZeroFlavor_Fritzsch2011JHEP,twoZeroFlavor_Zhou2016CPC}. Indeed, for example, one of the model parameters in Eq.(\ref{Eq:param_comlex}), such as one of the Majorana phases,  will be predicted by using the relation of $M_{e\tau} = M_{\mu\mu}=0$. We would like to discuss about the interplay between the scotogenic model and the two-zero textures of the flavor neutrino masses in the future work.
 
\section{Summary\label{sec:summary}}
There are two main topics in this paper. As the first topic, we have proposed a new parametrization of the complex Yukawa matrix in the scotogenic model  [Eq.(\ref{Eq:Y_Ye1Ye2Ye3}) with Eq.(\ref{Eq:a1a2a3a4a5a6})]. The new parametrization is compatible with the particle data group parametrization of the neutrino sector. Some analytical expressions for the neutrino masses in the scotogenic model with the new parametrization of the Yukawa matrix have been obtained. Although there are many other ways to parametrize the Yukawa matrix, the way in this paper may be one of the useful method to consider the phenomenology of the scotogenic model with the standard PDG parametrization.

As the second topic, we have considered some phenomenology of the socotogenic model with the one-zero-textures of neutrino flavor mass matrix. If the elements of the Yukawa matrix are real, one of the six patterns of the flavor neutrino mass matrix with single-zero element is favorable \cite{Kitabayashi2018PRD}. On the other hand, if the Yukawa matrix is complex, the five patterns ${\rm G_2}$, ${\rm G_3}$, ${\rm G_4}$, ${\rm G_5}$ and ${\rm G_6}$ are compatible with observations of the neutrino oscillations, dark matter relic abundance and branching ratio of the $\mu \rightarrow e\gamma$ process; however, ${\rm G_1}$ is unfavorable with observations.

Finally, we would like to comment that the baryon asymmetry of the universe is closely related to leptonic CP phases in the leptogenesis scenario \cite{Fukugita1986PLB}. The scenarios of leptogenesis in the scotogenic model have been extensively studied in the literature \cite{Ma2006MPLA,Suematsu2012EPJC,Kashiwase2012PRD,Kashiwase2012EPJC,Racker2014JCAP,Cai2017FrontPhys,Hugle2018PRD,Borah2018arXiv}. It seems that it is hard to realize ordinary thermal leptogenesis with flavor effects via the decay of the fermionic dark matter $N_i$ with a hierarchical mass spectrum \cite{Hugle2018PRD}. In this paper, we assume that the lightest Majorana singlet fermion is the dark matter with mass spectrum of $M_1 \le M_2 < M_3$. Thus, the vanilla leptogenesis scenario might not work properly for the model in this paper. This situation may be changed if we employ some alternative mechanism, such as resonant leptogenesis \cite{Pilaftsis1997PRD,Pilaftsis2004NPB}. Moreover, we know that in the bosonic dark matter scenario, the lightest neutral component in the scalar $\eta$ is a dark matter, a successful low-scale leptogenesis can be achieved \cite{Hugle2018PRD}. We would like to discuss the topics of the baryon asymmetry of the universe in the scotogenic model with the one-zero-textures of the neutrino flavor mass matrix as a separate work in the future.

\vspace{3mm}







\end{document}